\documentclass[journal,twoside]{IEEEtran}
\usepackage{amsfonts}
\usepackage{amssymb}
\usepackage{mathrsfs}
\usepackage{verbatim}
\usepackage{subfigure}
\usepackage[centertags]{amsmath}
\usepackage{graphicx}  
\newtheorem{thm}{\bf Theorem}
\newtheorem{cons}{ Construction}[section]

\newtheorem{defn}{Definition}

\newtheorem{eg}{Example}[section]

\ifCLASSINFOpdf
\else
\fi
\usepackage{array}

\begin{document}
%
\title{Algebraic Distributed Differential Space-Time Codes with Low Decoding Complexity}
%
%
%

\author{G. Susinder Rajan,~\IEEEmembership{Student Member,~IEEE,}
        and~B. Sundar~Rajan,~\IEEEmembership{Senior~Member,~IEEE}
\thanks{This work was supported through grants to B.S.~Rajan; partly by the IISc-DRDO program on Advanced Research in Mathematical Engineering, and partly by the Council of Scientific \& Industrial Research (CSIR, India) Research Grant (22(0365)/04/EMR-II). The material in this paper was presented in parts at the IEEE International Symposium on Information theory \cite{RaR4,RaR5} held at Nice, France during June 24-29, 2007. G. Susinder Rajan and B. Sundar Rajan are with the Department of Electrical Communication Engineering, Indian Institute of Science, Bangalore-560012, India. Email:\{susinder,bsrajan\}@ece.iisc.ernet.in.}
\thanks{Manuscript received May 07, 2007; revised September 10, 2007.}}

%
%

\markboth{IEEE Transactions on Wireless Communications ,~Vol.~xx, No.~xx, xxxx}{Rajan \MakeLowercase{and} Rajan: Algebraic Distributed Differential Space-Time Codes with Low Decoding Complexity}
%



\maketitle

\begin{abstract}
The differential encoding/decoding setup introduced by Kiran \emph{et al}, Oggier-Hassibi and Jing-Jafarkhani for wireless relay networks that use codebooks consisting of unitary matrices is extended to allow codebooks consisting of scaled unitary matrices. For such codebooks to be usable in the Jing-Hassibi protocol for cooperative diversity, the conditions involving the relay matrices and the codebook that need to be satisfied are identified. Using the algebraic framework of extended Clifford algebras, a new class of Distributed Differential Space-Time Codes satisfying these conditions for power of two number of relays and also achieving full cooperative diversity with a low complexity sub-optimal receiver is proposed. Simulation results indicate that the proposed codes outperform both the cyclic codes as well as the circulant codes. Furthermore, these codes can also be applied as Differential Space-Time codes for non-coherent communication in classical point to point multiple antenna systems.  
\end{abstract}

\begin{IEEEkeywords}
Algebra, cooperative diversity, low decoding complexity, space-time coding.
\end{IEEEkeywords}

%

\section{Introduction}
 \label{sec1}
%
%
%
%
\IEEEPARstart{C}{onsider} the scenario of $R+1$ users wanting to communicate to a single destination using cooperative strategies. For simplicity, we consider one of the $R+1$ users as the source and the remaining $R$ users as relays. Predominantly there are two types of relaying strategies discussed in the literature, i.e., 1) amplify and forward and 2) decode and forward. In this work, we focus only on amplify and forward based relaying strategies wherein the relays are allowed to perform linear processing of the received signal. Amplify and forward based relaying is simpler and moreover does not require the relays to have knowledge of the codebook used by the source or the knowledge of the source to relay channel fading gains. Recently in \cite{JiH1}, the idea of space-time coding for collocated MIMO (Multiple Input Multiple Output) systems has been applied in the setting of cooperative wireless relay networks in the name of distributing space-time coding, wherein coding is performed across users and time. This strategy provides each user a diversity order equal to the number of relays even though all the users are only equipped with a single antenna. The diversity thus achieved is called as cooperative diversity. However, such strategies require that the destination have complete knowledge of the fading coefficients from all the users to itself as well as that of the fading coefficients between users. Obtaining the knowledge of the fading coefficients between the users at the destination requires additional resources. To solve this problem, in \cite{KiR}, Kiran \emph{et al} have proposed a differential encoding/decoding setup for cooperative wireless networks that does not require the knowledge of fading coefficients between the users at the destination. Such codes were named as partially coherent distributed space-time codes in \cite{KiR}. In a recent work \cite{OgH1}, it has been shown that the same strategy of \cite{KiR} offers full diversity with a suboptimal receiver that does not require the knowledge of any of the fading coefficients. In an independent work \cite{JiJ}, Jing and Jafarkhani have proposed a differential encoding/decoding setup for cooperative wireless relay networks which is more general than the setup proposed in \cite{KiR,OgH1} and they have also provided few code constructions. We call the class of Differential Space-Time Codes (DSTCs) which can be used in a distributed manner for cooperative diversity as Distributed DSTCs (and denote by DDSTCs) to differentiate them from DSTCs for collocated MIMO systems. The problem of designing DDSTCs is more challenging than that of DSTCs for collocated MIMO systems, since in this scenario we have additional constraints to be satisfied which are due to the cooperative diversity protocol. In \cite{OgH2}, the setup of \cite{KiR,OgH1} has been generalized to the case when the source and destination nodes have multiple antennas. In \cite{OgH2}, the authors propose the use of a random diagonal unitary matrix codebook in general and for certain specific number of relays ($3$, $6$ and $9$) propose the use of cyclic codes. However, a random code is not guaranteed to offer full diversity. Moreover designing good cyclic codes becomes difficult for large number of relays. In \cite{JiJ}, the authors propose the use of Alamouti code and $Sp(2)$ code for the specific case of $2$ and $4$ relays. For an arbitrary number of relays, the authors construct fully diverse DDSTCs based on circulant matrices. But, except for the Alamouti code all other DDSTC constructions in the literature \cite{KiR,JiJ,OgH1,OgH2} have large decoding complexity. Thus, a general full diversity code construction targeting the requirements of low encoding complexity as well as low decoding complexity is not available. This issue gains significant importance especially if the number of cooperating terminals is large, which is quite expected in applications like wireless sensor networks. 

In this paper we address this issue and present the following contributions:
\begin{itemize}

\item The notion of encoding complexity for Space-Time codes in terms of $g$-group encodability is introduced (Definition \ref{defn1}) and its significance is highlighted.

\item The differential encoding/decoding setup introduced by Kiran \emph{et al} \cite{KiR}, Oggier \emph{et al} \cite{OgH2} and Jing \emph{et al} \cite{JiJ} for wireless relay networks that use codebooks consisting of unitary matrices is generalized to allow codebooks consisting of scaled unitary matrices. This generalization aids in reducing the encoding/decoding complexity of DDSTCs. When the codebook of scaled unitary matrices is used in the Jing-Hassibi protocol \cite{JiH1} for differential encoding at the source, the extra conditions involving the interrelationship between relay matrices and the codeword matrices that need to be satisfied are identified (Eqn. \eqref{eqn_extra_cond}).  

\item Using the algebraic framework of extended Clifford algebras, a new class of DDSTCs satisfying these extra conditions (Constructions \ref{cons_design} and \ref{cons_signalset}) is proposed which achieves full diversity with a low complexity sub-optimal receiver (Theorem \ref{thm_relay}). Explicit construction of multidimensional signal sets that lead to full diversity (Theorem \ref{thm_signalset}) are also provided. To the best of our knowledge, this is the first known low decoding complexity DDSTC scheme for cooperative wireless networks which is available for all power of two number of relays.
\end{itemize}

\subsection{Organization of the paper}
Section \ref{sec2} introduces the system model for the distributed differential encoding/decoding setup employing a differential scaled unitary matrix codebook at the source. In Section \ref{sec3}, the notion of $g$-group encodable and $g$-group decodable space-time codes is discussed for the system model of Section \ref{sec2} and the problem statement for DDSTCs with low encoding and decoding complexity is given. The extra conditions on the code structure imposed by the cooperative diversity protocol of \cite{KiR,JiJ,OgH1,OgH2} are then described. In Section \ref{sec4}, we explicitly construct a family of DSTCs using extended Clifford algebras and prove using algebraic techniques that they satisfy the extra conditions needed for DDSTCs. Simulation results and discussion on further work comprise Sections \ref{sec5} and \ref{sec6} respectively. 

\subsection{Notation}
For two sets $\mathscr{A}_1$ and $\mathscr{A}_2$, $\mathscr{A}_1\times\mathscr{A}_2$ denotes the Cartesian product of $\mathscr{A}_1$ and $\mathscr{A}_2$. For a matrix $\mathbf{C}$, the matrices $\mathbf{C}^*$, $\mathbf{C}^T$ and $\mathbf{C}^H$ denote the conjugate, transpose and conjugate transpose of $\mathbf{C}$ respectively. For a complex number $x$, $x_I$ and $x_Q$ denote its in-phase and quadrature-phase components respectively. For a vector $\mathbf{a}$, $\mathbf{a}[i]$ denotes the $i^{th}$ entry of $\mathbf{a}$. An all zero matrix of appropriate size will be denoted by $\mathbf{0}$. A complex Gaussian vector with zero mean and covariance matrix $\Omega$ will be denoted by $\mathcal{CN}(0,\Omega)$. 

\section{System model employing differential scaled unitary matrix codebook}
\label{sec2}  
In this section, we briefly explain the distributed differential encoding/decoding setup proposed in \cite{KiR,JiJ,OgH2} with a slight modification. 

We consider a network consisting of a source node, a destination node and $R$ other relay nodes which aid the source in communicating information to the destination. All the nodes are assumed to be equipped only with a single antenna and are half duplex constrained, i.e., a node cannot transmit and receive simultaneously at the same frequency. The wireless channels between the terminals are assumed to quasi-static and flat fading. The channel fading gains from the source to the $i$-th relay, $f_i$ and from the $j$-th relay to the destination $g_j$ are all assumed to be independent and identically distributed complex Gaussian random variables with zero mean and unit variance. Moreover, we assume symbol synchronization among all the nodes.

Every transmission cycle from the source to the destination comprises of two stages. In the first stage, the source transmits a $T(T\geq R)$ length vector $\sqrt{\pi_1TP}\mathbf{s}$ which the relays receive. Here, $P$ denotes the total average power spent by all the relays and the source. The fraction of total power $P$ spent by the source is denoted by $\pi_1$. The vector $\mathbf{s}$ satisfies $\mathrm{E}[\mathbf{s}^H\mathbf{s}]=1$ and is a function of the information that the source intends to communicate. The exact relation of $\mathbf{s}$ with the information sent by the source will be made precise in the sequel. The received vector at the $j$-th relay node is then given by $\mathbf{r_j}=\sqrt{\pi_1TP}f_j\mathbf{s}+\mathbf{v_j},\ \mathrm{with}\ \mathbf{v_j}\sim\mathcal{CN}(0,I_T)$. In the second half of the cycle, all the relay nodes are scheduled to transmit together. The $j$-th relay node transmits a $T$ length vector $\mathbf{t_j}$ which is a function of $\mathbf{r_j}$. The relays are only allowed to linearly process the received vector $\mathbf{r_j}$ or its conjugate $\mathbf{r_j}^*$. To be precise, the $j$-th relay node is equipped with a $T\times T$ unitary matrix $\mathbf{A_j}$ (called relay matrix) and it transmits either $\mathbf{t_j}=\sqrt{\frac{\pi_2P}{\pi_1P+1}}\mathbf{A_jr_j}$ or $\mathbf{t_j}=\sqrt{\frac{\pi_2P}{\pi_1P+1}}\mathbf{A_jr_j}^*$. Without loss of generality, we may assume that $M$ relays linearly process $\mathbf{r_j}$ and the remaining $R-M$ relays linearly process $\mathbf{r_j}^*$. Here, $\pi_2$ denotes the fraction of total power $P$ spent by a relay. If the quasi-static duration of the channel is much greater than $2T$ time slots, then the received vector at the destination is given by
\begin{equation}
\mathbf{y}=\sum_{j=1}^{R}g_j\mathbf{t_j}+\mathbf{w}=\sqrt{\frac{\pi_1\pi_2TP^2}{\pi_1P+1}}\mathbf{Xh}+\mathbf{n}
\end{equation}
where,
$$
\begin{array}{rl}
\mathbf{X}=&\left[\begin{array}{ccccccc}\mathbf{A_1s} & \dots & \mathbf{A_Ms} & \mathbf{A_{M+1}s}^* & \dots & \mathbf{A_Rs}^*\end{array}\right],\\
\mathbf{h}=&\left[\begin{array}{ccccccc}f_1g_1 & \dots f_Mg_M & f_{M+1}^*g_{M+1} & \dots & f_R^*g_R\end{array}\right]^T,\\
\mathbf{n}=&\sqrt{\frac{\pi_2P}{\pi_1P+1}}\left(\sum_{j=1}^{M}g_j\mathbf{A_jv_j}+\sum_{j=M+1}^{R}g_j\mathbf{A_jv_j}^*\right)+w
\end{array}
$$
and $\mathbf{w}\sim\mathcal{CN}(0,I_T)$ represents the additive noise at the destination. The power allocation factors $\pi_1$ and $\pi_2$ are chosen in such a way to satisfy $\pi_1P+\pi_2PR=2P$. Throughout this paper, we chose $\pi_1=1$ and $\pi_2=\frac{1}{R}$ as suggested in \cite{JiH1}.

The differential encoding is performed at the source as follows. An initial vector $\mathbf{s_0}$ known to the destination which satisfies $\mathbf{s_0}^H\mathbf{s_0}=1$ is transmitted by the source in the first cycle. The initial vector $s_0$ should be such that the initial matrix 
\begin{equation}
\label{eqn_initial_matrix}
\mathbf{X_0}=\left[\begin{array}{ccccccc}\mathbf{A_1s_0} & \dots & \mathbf{A_Ms_0} & \mathbf{A_{M+1}s_0}^* & \dots & \mathbf{A_Rs_0}^*\end{array}\right]
\end{equation}
is unitary. The transmitted vector at the $t$-th cycle is then given as follows
\begin{equation}
\label{eqn_diff_encoding}
\mathbf{s_t}=\frac{1}{a_{t-1}}\mathbf{U_ts_{t-1}}
\end{equation}
where, $\mathbf{U_t}\in\mathscr{C}$ is the codeword containing the information at the $t$-th cycle which satisfies $\mathbf{U_t}^H\mathbf{U_t}=a_t^2\mathbf{I_T},\ a_t\in\mathbb{R}$. The codebook $\mathscr{C}$ at the source consisting of scaled unitary matrices should satisfy $\mathrm{E}[a_t^2]=1$ in order to meet the power constraint at the source and relays. The originally proposed coding strategies in \cite{KiR,JiJ,OgH2} force $a_t=1$ for all codewords. The received vector at the destination in the $t$-th cycle can then be written as $\mathbf{y_t}=\sqrt{\frac{\pi_1\pi_2TP^2}{\pi_1P+1}}\mathbf{X_th_t}+\mathbf{n_t}$ where, \mbox{$\mathbf{X_t}=\left[\begin{array}{cccccc}\mathbf{A_1s_t} & \dots & \mathbf{A_Ms_t} & \mathbf{A_{M+1}s_t}^* & \dots & \mathbf{A_Rs_t}^*\end{array}\right]$}. If
\begin{equation}
\label{eqn_extra_cond}
\begin{array}{ccl}
\mathbf{A_iU_t}&=&\mathbf{U_tA_i},~\forall~\mathbf{U_t}\in\mathscr{C},~i=1,\dots,M~\mathrm{and}\\
~\mathbf{A_iU_t}^*&=&\mathbf{U_tA_i},~\forall~\mathbf{U_t}\in\mathscr{C},~i=M+1,\dots,R
\end{array}
\end{equation}
then, substituting for $\mathbf{s_t}$ from \eqref{eqn_diff_encoding} we get $\mathbf{X_t}=\frac{1}{a_{t-1}}\mathbf{U_tX_{t-1}}$. If the channel remains approximately constant for $4T$ channel uses, then we can assume that $\mathbf{h_t}=\mathbf{h_{t-1}}$. Then,
\begin{equation}
\begin{array}{rcl}
\mathbf{y_t}&=&\sqrt{\frac{\pi_1\pi_2TP^2}{\pi_1P+1}}\mathbf{X_th_t}+\mathbf{n_t}\\
&=&\sqrt{\frac{\pi_1\pi_2TP^2}{\pi_1P+1}}\frac{1}{a_{t-1}}\mathbf{U_tX_{t-1}h_{t-1}}+\mathbf{n_t}\\
&=&\frac{1}{a_{t-1}}\mathbf{U_ty_{t-1}}+\mathbf{\hat{n}_t}
\end{array}
\end{equation}
where, $\mathbf{\hat{n}_t}=-\frac{1}{a_{t-1}}\mathbf{U_tn_{t-1}}+\mathbf{n_t}$. Now we propose to decode the codeword $\mathbf{U_t}$ as follows
\begin{equation}
\label{eqn_metric0}
\mathbf{\hat{U_t}}=\mathrm{arg} \min_{\mathbf{U_t}\in\mathscr{C}}\parallel \mathbf{y_t}-\frac{1}{a_{t-1}}\mathbf{U_ty_{t-1}}\parallel^2
\end{equation}
where, $a_{t-1}$ can be estimated from the previous decision. Note that the above decoder does not require the knowledge of the channel gains and the relay matrices $\mathbf{A_i}, i=1,\dots,R$. Also, note that this decoder is not a Maximum-Likelihood (ML) decoder since the equivalent noise vector $\mathbf{\hat{n}_t}$ is dependent on the previous codeword $\mathbf{U_{t-1}}$ through $a_{t-1}$. However, for the collocated MIMO case, it has been shown in \cite{TaC,YGT2} that there is no significant performance loss due to this. In this setup, we call $\mathscr{C}$ a DDSTC to distinguish it from DSTCs for collocated MIMO systems. 

\section{Problem Statement}
\label{sec3}
In this section, we describe the problem statement of designing DDSTCs with low encoding complexity and low decoding complexity. 

In order to benefit from low encoding complexity and low decoding complexity, we propose to choose the DDSTC $\mathscr{C}$ to be a $g-$group encodable ($g>1$) linear STBC.
\begin{defn} 
\label{defn1}A linear design $\mathbf{S}(s_1,s_2,\dots,s_K)$ of size $n\times n$ in $K$ real indeterminates or variables $s_1,s_2,\dots,s_K$ is a $n\times n$ matrix with entries being a complex linear combination of the variables $s_1,s_2,\dots,s_K$. To be precise, it can be written as follows:
$$
\mathbf{S}(s_1,s_2,\dots,s_K)=\sum_{i=1}^{K}s_i\mathbf{B_i}
$$
where, $\mathbf{B_i}\in \mathbb C^{n\times n}$ are square complex matrices called the weight matrices. A linear STBC $\mathscr{C}$ is a finite set of $n\times n$ complex matrices which can be obtained by taking a linear design $\mathbf{S}(s_1,s_2,\dots,s_K)$ and specifying a signal set $\mathscr{A}\subset\mathbb{R}^{K}$ from which the information vector $\mathbf{x}=\left[\begin{array}{cccc}s_1 & s_2 & \dots & s_K\end{array}\right]^T$ take values, with the additional condition that $\mathbf{S}(\mathbf{a})\neq \mathbf{S}(\mathbf{a'}),~\forall~\mathbf{a}\neq \mathbf{a'}\in\mathscr{A}$. A linear STBC $\mathscr{C}=\left\{\mathbf{S}(\mathbf{x})|\mathbf{x}\in \mathscr{A}\right\}$ is said to be $g$-group encodable (or $\frac{K}{g}$ real symbol encodable or $\frac{K}{2g}$ complex symbol encodable) if $g$ divides $K$ and if $\mathscr{A}=\mathscr{A}_1\times\mathscr{A}_2\times\dots\times\mathscr{A}_g$ where each $\mathscr{A}_i,i=1,\dots,g\subset\mathbb{R}^{\frac{K}{g}}$.
\end{defn} 
Thus, given a $g$-group encodable STBC, it defines a natural partitioning of the real variables of its associated linear design into $g$-groups. This partitioning defines a partitioning of the set of weight matrices of $\mathbf{S}(\mathbf{x})$ into $g$-groups, the $k$-th group containing $K/g$ matrices. For simplicity we assume the simplest partitioning of the information symbol vector as $\mathbf{x}=\left[\begin{array}{cccc}\mathbf{x_1}^T & \mathbf{x_2}^T & \dots & \mathbf{x_g}^T\end{array}\right]^T$ where, \mbox{$\mathbf{x_k}=\left[\begin{array}{cccc}s_{\frac{(k-1)K}{g}+1} & s_{\frac{(k-1)K}{g}+1} & \dots & s_{\frac{kK}{g}}\end{array}\right]^T$}. Now $\mathbf{S}(\mathbf{x})$ can be written as,
$$
\mathbf{S}(\mathbf{x})=\sum_{k=1}^{g}\mathbf{S_k}(\mathbf{x_k})\quad\mathrm{where},~ \mathbf{S_k}(\mathbf{x_k})=\sum_{i=\frac{(k-1)K}{g}+1}^{\frac{kK}{g}}s_i\mathbf{B_i}.
$$
Minimizing the decoding metric corresponding to \eqref{eqn_metric0},
\begin{equation}
\label{eqn_metric}
\parallel \mathbf{y_t}-\frac{1}{a_{t-1}}\mathbf{S}(\mathbf{x})\mathbf{y_{t-1}}\parallel^2
\end{equation}
is in general not the same as minimizing
\begin{equation}
\label{eqn_submetric}
\parallel \mathbf{y_t}-\frac{1}{a_{t-1}}\mathbf{S_k}(\mathbf{x_k})\mathbf{y_{t-1}}\parallel^2
\end{equation}
for each $1\leq k\leq g$ individually. However if it so happens, then the decoding complexity is reduced by a large extent. 
\begin{defn}
\label{defn2}
A linear STBC $\mathscr{C}=\left\{\mathbf{S}(\mathbf{x})|\mathbf{x}\in \mathscr{A}\right\}$ is said to be $g$-group decodable (or $\frac{K}{g}$ real symbol decodable or $\frac{K}{2g}$ complex symbol decodable) if it is $g$-group encodable and if its decoding metric in \eqref{eqn_metric} can be simplified as in \eqref{eqn_submetric}.
\end{defn}
\begin{thm}\cite{RaR5}
The decoding metric in \eqref{eqn_metric} can be simplified as in \eqref{eqn_submetric} if 
\begin{equation}
\label{eqn_cond}
\mathbf{B_i}^H\mathbf{B_j}+\mathbf{B_j}^H\mathbf{B_i}=\mathbf{0}
\end{equation}
for all weight matrices $\mathbf{B_i}$ and $\mathbf{B_j}$ belonging to two different groups.
\end{thm}
We illustrate the significance of the above definitions concerning encoding and decoding complexity by giving few examples.
\begin{eg}
Consider the Golden code for $2$ transmit antennas. It has $8$ real variables. For the coherent MIMO channel, the signal set used is QAM for each complex variable. Hence, this code is a $8$-group encodable (since QAM is a Cartesian product of two PAM signal sets) and $1$-group decodable linear STBC. However, if we now impose the requirement for unitary codewords, then we will have to solve for signal sets which will yield unitary codewords inside the division algebra. This may amount to entangling all the $8$-real variables which will make the code $1$-group encodable and $1$-group decodable. This approach was recently attempted in \cite{Ogg}.
\end{eg}
\begin{eg}
\label{eg_alamouti}
Let us take the example of the Alamouti code for $2$ transmit antennas. It has $4$ real variables. If the signal set is chosen to be PSK for every complex variable, then all the codewords become unitary matrices. Hence the resulting code is $2$-group encodable and $2$-group decodable. However, note that if we take square QAM to be the signal set for each complex variable, then we get a $4$-group encodable (since square QAM is a Cartesian product of two PAM signal sets) and $4$-group decodable code, but now the codewords are scaled unitary matrices as opposed to unitary matrices. Thus relaxing the codewords to be scaled unitary matrices allows us to lower the encoding and decoding complexity.
\end{eg}
The above two examples show that {\it the choice of signal sets is crucial in obtaining low encoding and decoding complexity}. Example \ref{eg_alamouti} explicitly shows how allowing scaled unitary codebooks aids in reducing the encoding/decoding complexity. 

The DDSTC design problem is then to design a $g$-group decodable linear STBC 
$$
\mathscr{C}=\left\{\mathbf{S}\left(\mathbf{x}=\left[\begin{array}{cccc}s_1 & s_2 & \dots & s_K\end{array}\right]\right)|\mathbf{x}\in\mathscr{A}\right\}
$$
of size $T\times T$ to be used at the source such that

\begin{enumerate}
\item All codewords are scaled unitary matrices satisfying the transmit power constraint.
\item The parameters $K$ and $g$ are maximized. Increasing $K$ is motivated by the need for high rate transmission and maximizing $g$ is motivated by the requirement for low decoding complexity.
\item \label{cond_DDSTC4} There exist $R$ unitary matrices $\mathbf{A_1},\mathbf{A_2},\dots,\mathbf{A_R}$ of size $T\times T$ such that the first $M$ of them satisfy \mbox{$\mathbf{A_iC}=\mathbf{CA_i},~i=1,\dots,M,~\forall~ \mathbf{C}\in\mathscr{C}$} and the remaining $R-M$ of them satisfy \mbox{$\mathbf{A_iC}^*=\mathbf{CA_i},~i=M+1,\dots,R,~\forall~\mathbf{C}\in\mathscr{C}$}.
\item \label{cond_DDSTC5} There exists an initial vector $\mathbf{s_0}$ such that the initial matrix $\mathbf{X_0}$ is unitary.
\item $\min_{\mathbf{S_1},\mathbf{S_2}\in\mathscr{C}}|(\mathbf{S_1}-\mathbf{S_2})^H(\mathbf{S_1}-\mathbf{S_2})|$ is maximized, i.e., the coding gain is maximized.
\end{enumerate}

Observe that the requirements for designing DDSTCs are much more restrictive than that for DSTCs. Note that condition \eqref{cond_DDSTC4} and condition \eqref{cond_DDSTC5} are not required for designing DSTCs. As an additional requirement it would be nice to have a single design $\mathbf{S}(x_1,x_2,\dots,x_K)$ and a family of signal sets, one for each transmission rate such that all the required conditions are met. This means that we need to be able to find $R$ relay matrices satisfying the required conditions irrespective of the size of the code $|\mathscr{C}|$.

\section{Explicit construction of DDSTCs}
\label{sec4}

In this section we give an explicit construction of full diversity $4$-group decodable DDSTCs for all power of two number of relays. Optimizing the coding gain for the proposed DDSTCs is difficult in general and is beyond the scope of this paper. Constructing $g-$group decodable DDSTCs for arbitrary $g$ appears to be rather difficult and a brief discussion on the issues involved is given is Section \ref{sec6}. In Subsection \ref{subsection_design}, the construction of the proposed linear designs using tools from extended Clifford algebras is given. In Subsection \ref{subsection_signalset}, an explicit construction of signal sets $\mathscr{A}$ leading to full diversity for arbitrary transmission rate is provided for the linear designs given in Subsection \ref{subsection_design}, thus completely describing the construction of the linear STBC $\mathscr{C}$. In Subsection \ref{subsection_relay}, we use algebraic techniques to explicitly construct the $R$ relay matrices satisfying the required conditions for arbitrary size of the codebook $\mathscr{C}$.

\subsection{Construction of linear design}
\label{subsection_design}

In this subsection, we briefly describe a construction of a class of rate one, linear designs satisfying the conditions for four-group decodability which were first obtained using extended Clifford algebras in \cite{RaR2}. Extended Clifford algebras and linear designs from them are addressed in detail in \cite{RaR2}. However \cite{RaR2} chooses signal sets as applicable for coherent communication in a distributed space-time coding setup. A brief explanation of extended Clifford algebras and linear designs using them is provided in Appendix \ref{appendix1}. This algebraic framework given in Appendix \ref{appendix1} is needed to explicitly construct the relay matrices and the initial vector in Subsection \ref{subsection_relay}.

\begin{cons} 
Given a $n\times n$ linear design $\mathbf{A}(x_1,x_2,\dots,x_L)$ in $L$ complex variables $x_1,x_2,\dots,x_L$, one can construct a $2n\times 2n$ linear design $\mathbf{D}$ as follows:

$$
{\small
\mathbf{D}=\left[\begin{array}{cc}\mathbf{A}(x_1,x_2,\dots,x_L) & \mathbf{B}(x_{L+1},x_{L+2},\dots,x_{2L})\\\mathbf{B}(x_{L+1},x_{L+2},\dots,x_{2L}) & \mathbf{A}(x_1,x_2,\dots,x_L) \end{array}\right]
}
$$ 

\noindent where, the linear design $\mathbf{B}(x_{L+1},x_{L+2},\dots,x_{2L})$ is identical to the linear design $\mathbf{A}(x_1,x_2,\dots,x_L)$ except that it is a design in a different set of complex variables $x_{L+1},x_{L+2},\dots,x_{2L}$. We call this construction as the 'ABBA construction' which was first introduced in \cite{TBH}.
\end{cons}

\begin{cons} 
\label{cons_doubling}
Given a $n\times n$ linear design $\mathbf{A}(x_1,x_2,\dots,x_L)$, one can construct a $2n\times 2n$ linear design $\mathbf{S}$ as follows.  

$$
{\small
\mathbf{S}=\left[\begin{array}{lc}\mathbf{A}(x_1,x_2,\dots,x_L) & -\mathbf{B}^H(x_{L+1},x_{L+2},\dots,x_{2L})\\\mathbf{B}(x_{L+1},x_{L+2},\dots,x_{2L}) & \mathbf{A}^H(x_1,x_2,\dots,x_L) \end{array}\right]
}
$$

\noindent where, the linear design $\mathbf{B}(x_{L+1},x_{L+2},\dots,x_{2L})$ is identical to the linear design $\mathbf{A}(x_1,x_2,\dots,x_L)$. We call this construction as the 'doubling construction'. 
\end{cons}
We are now ready to describe the construction of the proposed linear designs. 
\begin{cons}
\label{cons_design}
To obtain a linear design for $R=2^\lambda$ relays, follow the steps given below.\\
Step 1: Starting with the linear design $[x_1]$ in complex variable $x_1$, keep applying ABBA construction iteratively on it till a $2^{\lambda-1}\times 2^{\lambda-1}$ linear design $\mathbf{D}$ is obtained.\\
Step 2: Apply doubling construction on $\mathbf{D}$ and scale it by $\frac{1}{\sqrt{R}}$ to obtain the linear design.
\end{cons}

\begin{eg}
Following the steps given above for $R=8$ relays, we get

\begin{equation} 
\label{eqn_design8}
\frac{1}{\sqrt{8}}\left[\begin{array}{ccccrrrr}
x_1 & x_2 & x_3 & x_4 & -x_5^* & -x_6^* & -x_7^* & -x_8^*\\
x_2 & x_1 & x_4 & x_3 & -x_6^* & -x_5^* & -x_8^* & -x_7^*\\ x_3 & x_4 & x_1 & x_2 & -x_7^* & -x_8^* & -x_5^* & -x_6^*\\ x_4 & x_3 & x_2 & x_1 & -x_8^* & -x_7^* & -x_6^* & -x_5^*\\ x_5 & x_6 & x_7 & x_8 & x_1^* & x_2^* & x_3^* & x_4^*\\ x_6 & x_5 & x_8 & x_7 & x_2^* & x_1^* & x_4^* & x_3^*\\ x_7 & x_8 & x_5 & x_6 & x_3^* & x_4^* & x_1^* & x_2^*\\
x_8 & x_7 & x_6 & x_5 & x_4^* & x_3^* & x_2^* & x_1^* 
\end{array}\right].
\end{equation} 
\end{eg}

\subsection{Construction of signal sets leading to full diversity}
\label{subsection_signalset}
In this subsection, we construct signal sets for the linear designs constructed in the previous subsection that lead to full diversity STBCs. In general, the signal sets should be designed such that the resulting linear STBC $\mathscr{C}$ meets the following requirements.
\begin{enumerate}
\item Scaled unitary codeword matrices meeting power constraint.
\item Four-group encodable and Four-group decodable.
\item Difference of any two different codeword matrices should be full rank. We call such a code to be 'fully diverse'.
\item The minimum determinant of the difference of any two codewords matrix should be maximized (Coding gain). 
\end{enumerate}
We shall first illustrate the signal set construction procedure for $4$ relays and derive important insights form it. Then, we generalize the ideas thus obtained for constructing signal sets for any $R=2^\lambda$ relays. The design for $4$ relays according to the construction in the previous subsection is given by 
\begin{equation}
\label{eqn_design4}
\mathbf{S}=\frac{1}{\sqrt{4}}\left[\begin{array}{ccrr}x_1 & x_2 & -x_3^* & -x_4^*\\
x_2 & x_1 & -x_4^* & -x_3^*\\
x_3 & x_4 & x_1^* & x_2^*\\
x_4 & x_3 & x_2^* & x_1^*\end{array}\right].
\end{equation}
Let us look at $\mathbf{S}^H\mathbf{S}$ shown in \eqref{eqn_long} at the top of the next page to find out the conditions on the signal sets under which the codewords are scaled unitary matrices.

\begin{figure*}
{\footnotesize
\begin{equation}
\label{eqn_long}
\mathbf{S}^H\mathbf{S}=\frac{1}{4}\left[\begin{array}{cccc}
\sum_{i=1}^{4}\vert x_i\vert^2 & x_1^*x_2+x_2^*x_1+x_3^*x_4+x_4^*x_3 & 0 & 0\\
x_1^*x_2+x_2^*x_1+x_3^*x_4+x_4^*x_3 & \sum_{i=1}^{4}\vert x_i\vert^2 & 0 & 0\\
0 & 0 & \sum_{i=1}^{4}\vert x_i\vert^2 & x_1^*x_2+x_2^*x_1+x_3^*x_4+x_4^*x_3\\
0 & 0 & x_1^*x_2+x_2^*x_1+x_3^*x_4+x_4^*x_3 & \sum_{i=1}^{4}\vert x_i\vert^2\end{array}\right] 
\end{equation}
}
\hrule
\end{figure*}

From \eqref{eqn_long}, we see that the signal set should be chosen such that the following condition is satisfied for all the signal points.
\begin{equation}
\label{eqn_non_orth}
x_1^*x_2+x_2^*x_1+x_3^*x_4+x_4^*x_3=0.
\end{equation}
Firstly, we identify the grouping of the real variables into groups such that the corresponding weight matrices in each group satisfy \eqref{eqn_cond}. According to the construction, the four groups or real variables are as follows: First group: $\left\{x_{1I},x_{2I}\right\}$, Second group: $\left\{x_{1Q},x_{2Q}\right\}$, Third group: $\left\{x_{3I},x_{4I}\right\}$, Fourth group: $\left\{x_{3Q},x_{4Q}\right\}$. It is important to choose signal sets such that they do not enforce joint constraints on variables from different groups. The requirement for scaled unitary codewords as in \eqref{eqn_non_orth} can be satisfied without disturbing $4$-group encodability if all the signal points satisfy the following equation: 
\begin{equation}
\label{eqn_signalset1}
x_{1I}x_{2I}=-x_{1Q}x_{2Q}=c_1,\quad
x_{3I}x_{4I}=-x_{3Q}x_{4Q}=c_2
\end{equation}
where, $c_1$ and $c_2$ are positive real constants. Then, the average power constraint requirement can be met by satisfying the conditions,
\begin{equation}
\label{eqn_signalset2}
\begin{array}{c}
\mathrm{E}(x_{1I}^2+x_{2I}^2)=1,~\mathrm{E}(x_{1Q}^2+x_{2Q}^2)=1,\\
\mathrm{E}(x_{3I}^2+x_{4I}^2)=1,~\mathrm{E}(x_{3Q}^2+x_{4Q}^2)=1.
\end{array}
\end{equation}
A common set of solutions for \eqref{eqn_signalset1} and \eqref{eqn_signalset2} can be obtained by taking points on the intersection of circles and hyperbolas. To meet the third requirement of full diversity we make use of the structure of the constructed designs. Note that 

\begin{equation}
\label{eqn_commute}
{\scriptsize
\begin{array}{rl}
|\Delta \mathbf{S}^H\Delta \mathbf{S}|&=\frac{1}{4}\left|\left[\begin{array}{cc}\Delta \mathbf{A}^H\Delta \mathbf{A}+\Delta \mathbf{B}^H\Delta \mathbf{B} & (\Delta \mathbf{A}\Delta \mathbf{B}-\Delta \mathbf{B}\Delta \mathbf{A})^H\\
\Delta \mathbf{A}\Delta \mathbf{B}-\Delta \mathbf{B}\Delta \mathbf{A} & \Delta \mathbf{A}\Delta \mathbf{A}^H+\Delta \mathbf{B}\Delta \mathbf{B}^H\end{array}\right]\right|\\
\\
&=\frac{1}{4}\left|\left[\begin{array}{cc}\Delta \mathbf{A}^H\Delta \mathbf{A}+\Delta \mathbf{B}^H\Delta \mathbf{B} & \mathbf{0}\\
\mathbf{0} & \Delta \mathbf{A}\Delta \mathbf{A}^H+\Delta\mathbf{B}\Delta \mathbf{B}^H\end{array}\right]\right|\\
&\geq \frac{1}{4}\mathrm{max}(|\Delta \mathbf{A}|^*|\Delta \mathbf{A}|,|\Delta \mathbf{B}|^*|\Delta \mathbf{B}|)^2
\end{array}
}
\end{equation}

\noindent where, the notation $\Delta$ has been used to denote the difference matrix which will have the same form as the associated linear design. The second equality in \eqref{eqn_commute} follows because the linear designs $\mathbf{A}$ and $\mathbf{B}$ commute. This can be proved as follows. Let $\mathbf{A}=\left[\begin{array}{cc}\mathbf{W} & \mathbf{X}\\\mathbf{X} & \mathbf{W} \end{array}\right]$ and \mbox{$\mathbf{B}=\left[\begin{array}{cc}\mathbf{W'} & \mathbf{X'}\\\mathbf{X'} & \mathbf{W'} \end{array}\right]$}. Then, we have

$$
{\scriptsize
\begin{array}{l}
\left[\begin{array}{cc}\mathbf{W} & \mathbf{X}\\\mathbf{X} & \mathbf{W} \end{array}\right]\left[\begin{array}{cc}\mathbf{W'} & \mathbf{X'}\\\mathbf{X'} & \mathbf{W'} \end{array}\right]=\left[\begin{array}{cc}\mathbf{WW'}+\mathbf{XX'} & \mathbf{WX'+XW'}\\\mathbf{XW'+WX'} & \mathbf{XX'+WW'} \end{array}\right],\\
\\
\left[\begin{array}{cc}\mathbf{W'} & \mathbf{X'}\\\mathbf{X'} & \mathbf{W'} \end{array}\right]\left[\begin{array}{cc}\mathbf{W} & \mathbf{X}\\\mathbf{X} & \mathbf{W} \end{array}\right]=\left[\begin{array}{cc}\mathbf{W'W}+\mathbf{X'X} & \mathbf{W'X}+\mathbf{X'W}\\\mathbf{X'W}+\mathbf{W'X} & \mathbf{X'X}+\mathbf{W'W} \end{array}\right].
\end{array}
}
$$

Thus we observe that linear designs $\mathbf{A}$ and $\mathbf{B}$ commute if their constituent sub-matrices commute. Applying this argument recursively, the claim follows.

Thus $\mathbf{S}$ is fully diverse, i.e., $|\Delta \mathbf{S}|\neq 0$ if either of the submatrices $\mathbf{A}$ or $\mathbf{B}$ is fully diverse. For the example of $4$ relays, we have \mbox{$|\Delta \mathbf{A}|=\left|\left[\begin{array}{cc}\Delta x_1 & \Delta x_2\\ \Delta x_2 & \Delta x_1 \end{array}\right]\right|=(\Delta x_1+\Delta x_2)(\Delta x_1-\Delta x_2)$}. Hence we can guarantee full diversity by ensuring that $\Delta x_1\neq\pm\Delta x_2$ and $\Delta x_3\neq\pm\Delta x_4$. Just like before, we should be careful not to disturb $4$-group encodability in the process. We take care of that requirement also by satisfying the following conditions:

\begin{equation}
\label{eqn_fulldiversity}
\begin{array}{c}
\Delta x_{1I}\neq\pm\Delta x_{2I},~\Delta x_{1Q}\neq\pm\Delta x_{2Q},\\
\Delta x_{3I}\neq\pm\Delta x_{4I},~\Delta x_{3Q}\neq\pm\Delta x_{4Q}.
\end{array}
\end{equation}

\begin{figure}[h]
\centering
\includegraphics[width=3.5in]{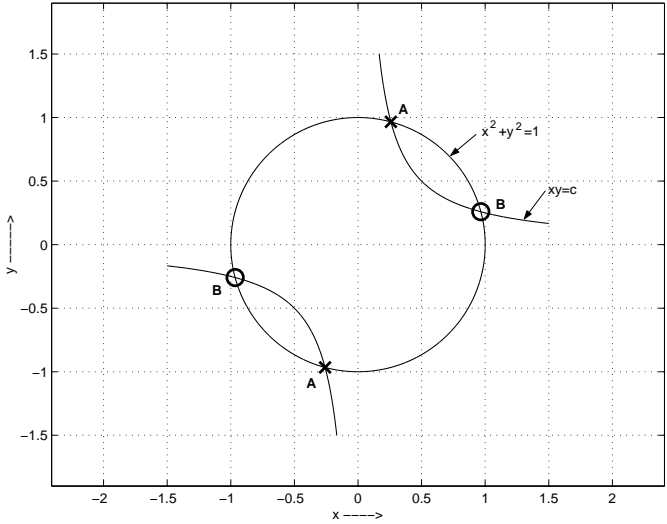}
\caption{Signal set structure in 2 dimensions}
\label{fig_signal2dimstruct}
\end{figure} 

Putting together all the conditions that need to be met, we have to choose constellation points satisfying \eqref{eqn_signalset1}, \eqref{eqn_signalset2} and \eqref{eqn_fulldiversity}. The solution can be found simply by finding the intersection of points on the unit circle $x^2+y^2=1$ with a hyperbola $xy=c$, where $c<1$ on the two dimensional $xy$ plane. This is illustrated in Fig. \ref{fig_signal2dimstruct}.

\begin{figure*}
\centering
\includegraphics{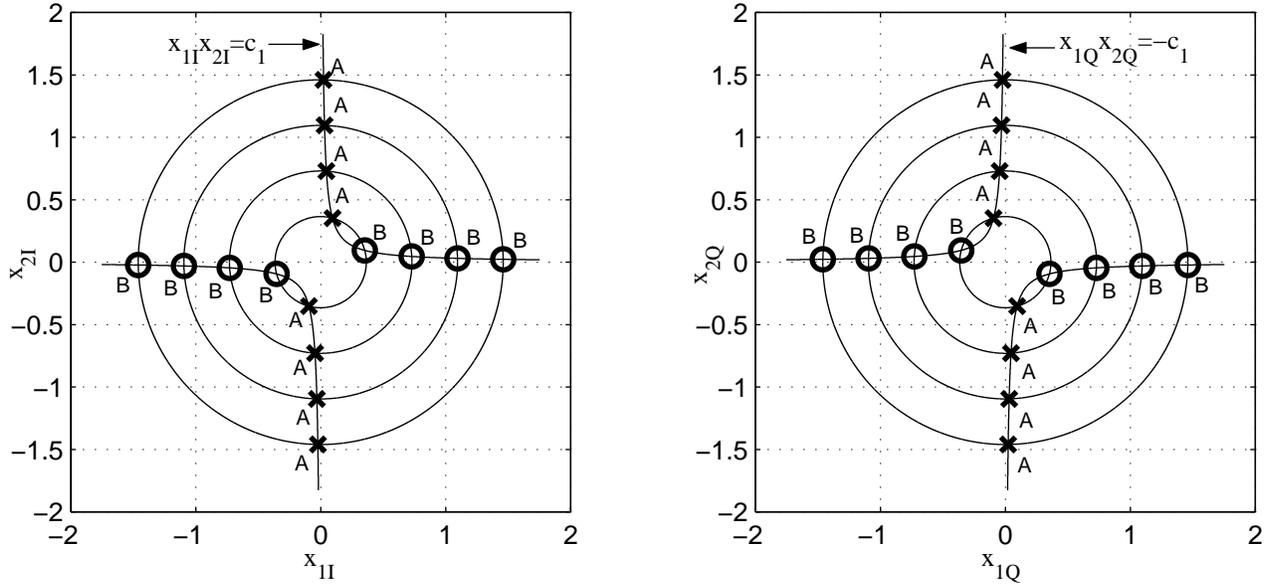}
\caption{General signal set for four relays}
\label{fig_signal2dim}
\end{figure*}

Observe from Fig. \ref{fig_signal2dimstruct} that the hyperbola intersects the circle at four different points. But the full diversity criterion demands that $\Delta x\neq\pm\Delta y$. After enforcing this condition, only two points survive out of the four points. They can be either the set of points marked A or the set of points marked B in Fig. \ref{fig_signal2dimstruct}. Thus we have obtained a signal set containing $2$ points. Supposing that we need more than two points, we can draw more circles (centered at origin) with radii such that the average power constraint is met and then find those points intersecting with the hyperbola. More precisely, to get $m$ points, we draw $\frac{m}{2}$ concentric circles with increasing radii $r_1,r_2,\dots,r_{\frac{m}{2}}$ such that $\sum_{i=1}^{\frac{m}{2}}r_i^2=\frac{m}{2}$. Then we find those points intersecting with the hyperbola $xy=c$ where, $c$ is a positive number less than\footnote{This condition is necessary since otherwise the hyperbola will not intersect the circle with least radius.} $r_1^2$. In this manner we can get the desired signal set for the variables $x_{1I},x_{2I}$ and $x_{3I},x_{4I}$. The signal set for the variables $x_{1Q},x_{2Q}$ and $x_{3Q},x_{4Q}$ can be obtained by considering a different hyperbola. This is illustrated in Fig. \ref{fig_signal2dim} shown at the top of the next page.

\begin{figure}[h]
\centering
\includegraphics[width=3.5in]{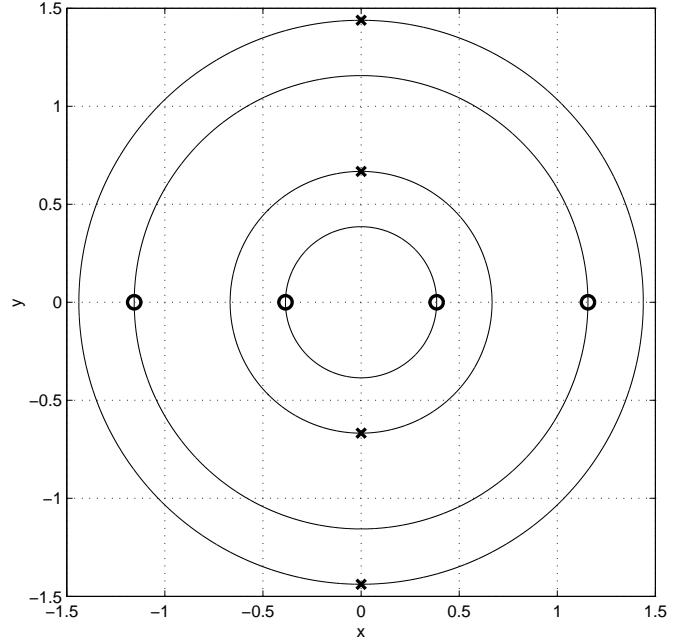}
\caption{Proposed signal set in 2 dimensions}
\label{fig_signal2dim2}
\end{figure}

Based on empirical studies, we propose to choose points in the set \textbf{A} and set \textbf{B} alternatively on each circle. Also, choosing the constants $c_1=c_2=0$ seems to give the largest coding gain. Similar observations have also been made in \cite{YGT2} for another class of linear STBCs applicable in co-located MIMO systems. Implementing these two suggestions makes the signal identical for all the $4$-groups and the resulting signal set for the case of $8$ points is shown in Fig.\ref{fig_signal2dim2}. Note that the proposed signal set construction is far from general and the most general solution would correspond to the following equations:
$$
\begin{array}{c}
x_{1I}x_{2I}=-x_{1Q}x_{2Q}=c_1,~x_{3I}x_{4I}=-x_{3Q}x_{4Q}=c_2,\\
\mathrm{E}(x_{1I}^2+x_{2I}^2)=d_1,~\mathrm{E}(x_{1Q}^2+x_{2Q}^2)=d_2,\\
\mathrm{E}(x_{3I}^2+x_{4I}^2)=d_3,~\mathrm{E}(x_{3Q}^2+x_{4Q}^2)=d_4,\\
\Delta x_{1I}\neq\pm\Delta x_{2I},~\Delta x_{1Q}\neq\pm\Delta x_{2Q},\\
\Delta x_{3I}\neq\pm\Delta x_{4I},~\Delta x_{3Q}\neq\pm\Delta x_{4Q}.
\end{array}
$$

In general, we need to choose the parameters $c_1,c_2,d_1,d_2,d_3,d_4,r_1,r_2,\dots,r_{\frac{m}{2}}$ and the solution points such that the coding gain is maximized. 

Based on the detailed illustration of signal set construction for $4$ relays antennas, a natural generalization of it to higher dimensions is given in Construction \ref{cons_signalset} as follows.

\begin{cons}
\label{cons_signalset}
Suppose that we want a $Q$ points signal set$\subset\mathbb{R}^{2^{\lambda+1}}$ for the constructed design for $R=2^\lambda$ relays. Then the resulting signal set $\subset\mathbb{R}^{2^{\lambda+1}}$ should be a Cartesian product of $4$ signal sets in $\mathbb{R}^{{2^{\lambda-1}}}$ since we insist on $4$-group encodability. In our case, we choose all the four sets to be identical and each will contain $\sqrt[4]{Q}$ points. Let the signal points in $\mathbb{R}^{{2^{\lambda-1}}}$ be labeled as $\mathbf{p}_i,\ i=1,\dots,\sqrt[4]{Q}$. If $i=2q+r$, for some integers $q$ and $r$ where $1\leq r \leq 2$, then $\mathbf{p}_i$ is given by
\begin{equation}
\begin{array}{c}
\mathbf{p}_i[j]=0\ \forall j\neq(q\ \mathrm{mod}\ 2^{\lambda-1})+1\\
\mathbf{p}_i[(q\ \mathrm{mod}\ 2^{\lambda-1})+1]=+r_{q+1},\ \mathrm{if}\ r=1\\
\mathbf{p}_i[(q\ \mathrm{mod}\ 2^{\lambda-1})+1]=-r_{q+1},\ \mathrm{if}\ r=2
\end{array}
\end{equation} 
where, $r_i,\ i=1,\dots,\frac{\sqrt[4]{Q}}{2}$ are positive real numbers such that $r_{i+1}>r_{i},\ \forall i=1,\dots,\frac{\sqrt[4]{Q}}{2}-1$ and $\sum_{i=1}^{\frac{\sqrt[4]{Q}}{2}}r_i^2=\frac{\sqrt[4]{Q}}{2}$.
\end{cons} 

\begin{eg}
Let $R=2^3=8$ and $Q=16^4$. Thus the rate of transmission of this code will be $\frac{\log_2 Q}{8}=2$ bits per channel use. The design is given in \eqref{eqn_design8}. The corresponding four dimensional signal set is described below.

\begin{equation}
\begin{array}{c}
\mathbf{p}_1=\left[\begin{array}{cccc}r_1 & 0 & 0 & 0\end{array}\right]^T,~ \mathbf{p}_2=\left[\begin{array}{cccc}-r_1 & 0 & 0 & 0\end{array}\right]^T,\\
\mathbf{p}_3=\left[\begin{array}{cccc}0 & r_2 & 0 & 0\end{array}\right]^T,~
\mathbf{p}_4=\left[\begin{array}{cccc}0 & -r_2 & 0 & 0\end{array}\right]^T,\\
\mathbf{p}_5=\left[\begin{array}{cccc}0 & 0 & r_3 & 0\end{array}\right]^T,~
\mathbf{p}_6=\left[\begin{array}{cccc}0 & 0 & -r_3 & 0\end{array}\right]^T,\\
\mathbf{p}_7=\left[\begin{array}{cccc}0 & 0 & 0 & r_4\end{array}\right]^T,~
\mathbf{p}_8=\left[\begin{array}{cccc}0 & 0 & 0 & -r_4\end{array}\right]^T,\\
\mathbf{p}_9=\left[\begin{array}{cccc}r_5 & 0 & 0 & 0\end{array}\right]^T,~
\mathbf{p}_{10}=\left[\begin{array}{cccc}-r_5 & 0 & 0 & 0\end{array}\right]^T,\\
\mathbf{p}_{11}=\left[\begin{array}{cccc}0 & r_6 & 0 & 0\end{array}\right]^T,~
\mathbf{p}_{12}=\left[\begin{array}{cccc}0 & -r_6 & 0 & 0\end{array}\right]^T,\\
\mathbf{p}_{13}=\left[\begin{array}{cccc}0 & 0 & r_7 & 0\end{array}\right]^T,~
\mathbf{p}_{14}=\left[\begin{array}{cccc}0 & 0 & -r_7 & 0\end{array}\right]^T,\\
\mathbf{p}_{15}=\left[\begin{array}{cccc}0 & 0 & 0 & r_8\end{array}\right]^T,~
\mathbf{p}_{16}=\left[\begin{array}{cccc}0 & 0 & 0 & -r_8\end{array}\right]^T
\end{array}
\end{equation}

\noindent where, $r_1=0.3235$, $r_2=\sqrt{3}r_1$, $r_5=3r_1$, $r_3=r_2+\frac{r_5-r_2}{3}$, $r_4=r_2+2\frac{r_5-r_2}{3}$, $r_6=\left(2+\sqrt{3}\right)r_1$, $r_7=r_3+2r_1$, $r_8=r_4+2r_1$. Though the dimension of the signal set is $4$, due to the special structure of the signal set, we can study the two dimensional projections of the signal points which is graphically shown in Fig. \ref{fig_signal4dim} at the top of the next page.

\begin{figure*}
\centering
\includegraphics{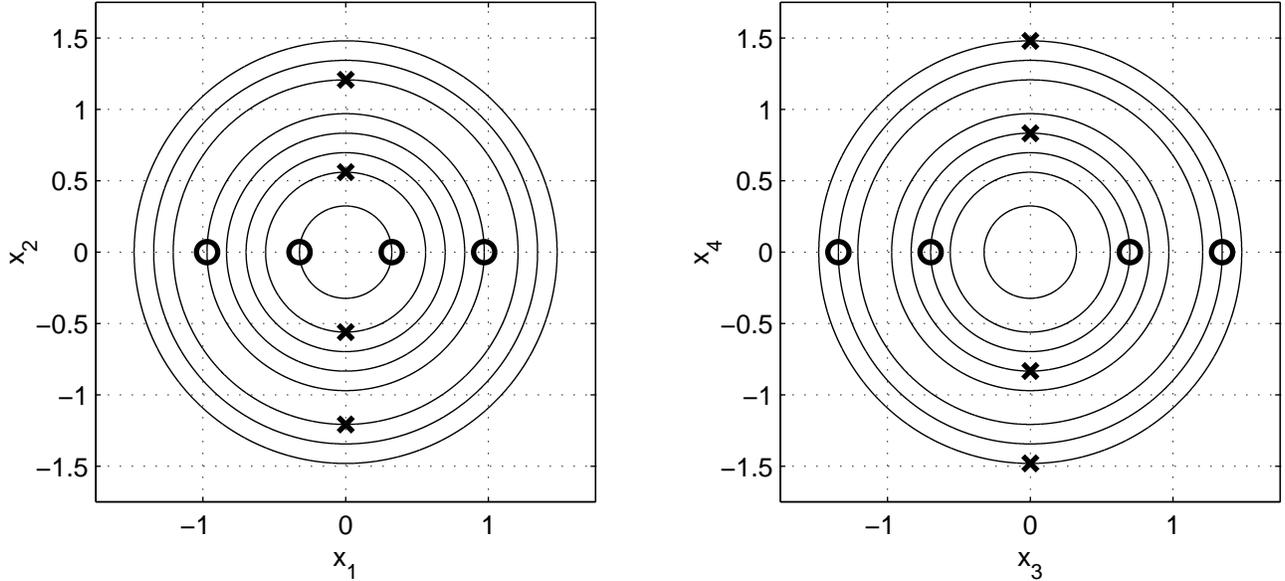}
\caption{Two dimensional projections of four dimensional signal set for Eight relays}
\label{fig_signal4dim}
\end{figure*} 
\end{eg}

The proposed multidimensional signal sets in Construction \ref{cons_signalset} are far from generality and obtaining a general solution targeting to maximize the coding gain appears to be difficult.

\begin{thm}
\label{thm_signalset}
Construction \ref{cons_signalset} provides fully diverse signals sets for the designs given in Construction \ref{cons_design}.  
\end{thm}
\begin{IEEEproof}
The constructed designs for $2L$ complex variables has the structure as shown in Construction \ref{cons_doubling}. The partitioning of the real variables into four groups is as follows.
\begin{enumerate}
\item First group : $\left\{x_{1I},x_{2I},\dots,x_{LI}\right\}$
\item Second group : $\left\{x_{1Q},x_{2Q},\dots,x_{LQ}\right\}$
\item Third group : $\left\{x_{(L+1)I},x_{(L+2)I},\dots,x_{2LI}\right\}$
\item Fourth group : $\left\{x_{(L+1)Q},x_{(L+2)Q},\dots,x_{2LQ}\right\}$
\end{enumerate}
The signal set is a Cartesian product of four smaller dimensional signal sets $\subset \mathbb{R}^L$. Since \mbox{$\mathbf{S}^H\mathbf{S}=\left[\begin{array}{cc}\mathbf{A}^H\mathbf{A}+\mathbf{B}^H\mathbf{B} & \mathbf{0}\\
\mathbf{0} & \mathbf{A}\mathbf{A}^H+\mathbf{B}\mathbf{B}^H\end{array}\right]$} and because the designs $\mathbf{A}$ and $\mathbf{B}$ are identical, it is sufficient to show that the design $\mathbf{A}(x_1,x_2,\dots,x_L)$ gives rise to fully diverse scaled unitary codewords for the chosen signal set. Let the given signal set be denoted by $\Lambda\subset\mathbb{R}^L$. Then the following properties are satisfied $\forall~ x,y\in\Lambda,~x\neq y$.
\begin{eqnarray}
\label{eqn_scaledunitarycond}
x[i]x[j]&=&0,~ 1\leq i\neq j\leq L\\
\label{eqn_fulldiv1}
1\leq|\left\{i|~(x-y)[i]\neq 0 \right\}|&\leq&2\\
\label{eqn_fulldiv2}
(x-y)[i]&\neq&\pm(x-y)[j],~ i\neq j
\end{eqnarray}
The design $\mathbf{A}(x_1,x_2,\dots,x_L)$ can be obtained by repeatedly applying ABBA construction on the linear design $[x_1]$. The off-diagonal entries of the matrix $\mathbf{A}(x_1,x_2,\dots,x_L)^H\mathbf{A}(x_1,x_2,\dots,x_L)$ are sum of terms like $x_i^*x_j+x_j^*x_i,~i\neq j$. But we have, \mbox{$x_i^*x_j+x_j^*x_i=2\left(x_{iI}x_{jI}+x_{iQ}x_{jQ}\right)$} which is equal to zero for the signal set $\Lambda$ by virtue of \eqref{eqn_scaledunitarycond}. Thus we are guaranteed of scaled unitary codewords. 

Due to the ABBA structure of the linear design $\mathbf{A}(x_1,x_2,\dots,x_L)$ , the determinant can be obtained in an iterative manner. Let $\mathbf{A}(x_1,x_2,\dots,x_K)=\left[\begin{array}{cc}\mathbf{A_1} & \mathbf{A_2}\\\mathbf{A_2} & \mathbf{A_1}\end{array}\right]$. Then we have

\begin{equation*}
\begin{array}{rcl}
|\mathbf{A}|&=&\left|\left[\begin{array}{cc}\mathbf{I} & \mathbf{0}\\\mathbf{I} & \mathbf{I}\end{array}\right]\left[\begin{array}{cc}\mathbf{A_1} & \mathbf{A_2}\\\mathbf{A_2} & \mathbf{A_1}\end{array}\right]\left[\begin{array}{cc}\mathbf{I} & \mathbf{0}\\-\mathbf{I} & \mathbf{I}\end{array}\right]\right|\\
\\
&=&\left|\left[\begin{array}{cc}\mathbf{A_1}-\mathbf{A_2} & \mathbf{A_2}\\\mathbf{0} & \mathbf{A_1}+\mathbf{A_2}\end{array}\right]\right|\\
&=&\left(|\mathbf{A_1}+\mathbf{A_2}|\right)\left(|\mathbf{A_1}-\mathbf{A_2}|\right).
\end{array}
\end{equation*}

The above equation suggests that $|\mathbf{A}(x_1,x_2,\dots,x_L)|$ can be computed recursively. For $L=2$, \mbox{$|\mathbf{A}(x_1,x_2)|=(x_1+x_2)(x_1-x_2)$}. Using this, we can easily show that $|\mathbf{A}(x_1,x_2,\dots,x_L)|$ is a product of $L$ terms. A typical term in the product looks like $\left(x_1\pm x_2\pm\dots\pm x_L\right)$. Thus to ensure $|\Delta \mathbf{A}(x_1,x_2,\dots,x_L)|\neq 0$, each term in the product should not equal zero. Thus we need \mbox{$\left(\Delta x_1\pm \Delta x_2\pm\dots\pm \Delta x_L\right)\neq 0$}. Let us look at the real part of this expression. We get $\left(\Delta x_{1I}\pm \Delta x_{2I}\pm\dots\pm \Delta x_{LI}\right)$. For the given signal set $\Lambda$, from \eqref{eqn_fulldiv1} and \eqref{eqn_fulldiv2}, we have \mbox{$\left(\Delta x_{1I}\pm \Delta x_{2I}\pm\dots\pm \Delta x_{LI}\right)\neq 0$}. Thus the theorem is proved.
\end{IEEEproof}


\subsection{Construction of Relay Matrices and Initial vector}
\label{subsection_relay}

\begin{thm} 
\label{thm_relay}
For the linear designs given by Construction \ref{cons_design}, there exist $R$ relay matrices satisfying condition \eqref{cond_DDSTC4} for arbitrary signal sets. Moreover, if the initial vector \mbox{$\mathbf{s_0}=\left[\begin{array}{cccc}1 & 0 & \dots & 0\end{array}\right]$}, then the initial matrix $\mathbf{X_0}$ given in \eqref{eqn_initial_matrix} becomes unitary.
\end{thm}
\begin{IEEEproof} 
See Appendix \ref{appendix2}.
\end{IEEEproof}

\begin{eg}
\label{eg_4relay}
Let $R=4$. Then the DDSTC $\mathscr{C}$ is obtained using the design shown in \eqref{eqn_design4} and the signal set given in Construction \ref{cons_signalset}. The signal set is a Cartesian product of four $2$-dimensional signal sets. The relay matrices are given as follows:
$$
\begin{array}{c}
\mathbf{A_1}=\mathbf{I_4},~\mathbf{A_2}=\left[\begin{array}{cccc}0 & 1 & 0 & 0\\
1 & 0 & 0 & 0\\
0 & 0 & 0 & 1\\
0 & 0 & 1 & 0\end{array}\right],\\
\mathbf{A_3}=\left[\begin{array}{ccrr}0 & 0 & -1 & 0\\
0 & 0 & 0 & -1\\
1 & 0 & 0 & 0\\
0 & 1 & 0 & 0\end{array}\right],~\mathbf{A_4}=\left[\begin{array}{ccrr}0 & 0 & 0 & -1\\
0 & 0 & -1 & 0\\
0 & 1 & 0 & 0\\
1 & 0 & 0 & 0\end{array}\right].
\end{array}
$$
The initial vector $\mathbf{s_0}=\left[\begin{array}{cccc}1 & 0 & \dots & 0\end{array}\right]^T$ and the initial matrix $\mathbf{X_0}=\mathbf{I_4}$. This DDSTC is single complex symbol decodable (or 2 real symbol decodable).
\end{eg}

\section{Simulation Results}
\label{sec5}
In this section, we compare the error performance of the proposed DDSTC in Example \ref{eg_4relay} with that of the circulant codes in \cite{JiJ} and the cyclic codes in \cite{OgH2} for transmission rates of $1$ bit per channel use (bpcu) and $1.5$ bpcu. The signal set chosen for the proposed code is as given by Construction $\ref{cons_signalset}$ with parameters $r_1=\frac{1}{\sqrt{3}}$, $r_2=\sqrt{\frac{5}{3}}$ for a transmission rate of $1$ bpcu and for the case of rate=1.5 bpcu, the signal set parameters are $r_1=0.378$, $r_2=0.8452$, $r_3=1.1339$ and $r_4=1.3628$. The circulant code of \cite{JiJ} is given by $\left\{u_1\mathbf{A_1},u_2\mathbf{A_2},u_3\mathbf{A_3},u_4\mathbf{A_4}|u_i\in\mathscr{F}_i,i=1,\dots,4\right\}$. The relay matrices $\mathbf{A_i}$ are taken to be powers of the $4\times 4$ circulant matrix. The signal set $\mathscr{F}_i$ is chosen to be $\theta_i$ radians rotated version of $64$-PSK for a transmission rate of $1$ bpcu and $1024$-PSK for a rate of $1.5$ bpcu. The rotation angles are chosen as follows to guarantee full diversity: $\theta_1=0$, $\theta_2=1.5$, $\theta_3=3$, $\theta_4=4.5$. Another reference for comparison is the cyclic codes proposed in \cite{OgH2} whose codewords for rate=$1$ bpcu and rate=$1.5$ bpcu are given as follows:\\

$\left\{\left[\begin{array}{cccc}
\zeta_{256} & 0 & 0 & 0\\
0 & \zeta_{256}^{11} & 0 & 0\\
0 & 0 & \zeta_{256}^{67} & 0\\
0 & 0 & 0 & \zeta_{256}^{101}
\end{array}
\right]^i, i=0,\dots,255\right\}$ and $\left\{\left[\begin{array}{cccc}
\zeta_{4096} & 0 & 0 & 0\\
0 & \zeta_{4096}^{43} & 0 & 0\\
0 & 0 & \zeta_{4096}^{877} & 0\\
0 & 0 & 0 & \zeta_{4096}^{2039}
\end{array}
\right]^i, i=0,\dots, 4095\right\}$ where, \mbox{$\zeta_{256}=e^{\frac{2\pi i}{256}}$} and $\zeta_{4096}=e^{\frac{2\pi i}{4096}}$. For simulations, we have assumed a block fading channel which is quasi-static for $800$ channel uses and varies independently from one block to another. 

\begin{figure}[h]
\centering
\includegraphics[width=3.5in]{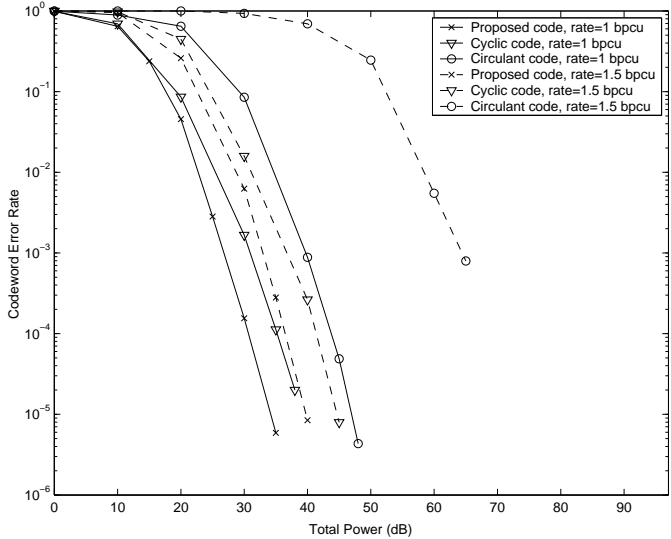}
\caption{Error performance comparison of the proposed codes with those of \cite{JiJ} and \cite{OgH2}}
\label{fig_simulation}
\end{figure} 

\mbox{Fig. \ref{fig_simulation}} shows the error performance curves of the proposed codes in comparison with those of \cite{JiJ,OgH2}. It can be observed from Fig. \ref{fig_simulation} that for a transmission rate of $1$ bpcu, the proposed code outperforms the cyclic code by about $5$ dB and the circulant code by more than $10$ dB. Similarly for a transmission rate of $1.5$ bpcu, it can be observed that the proposed code outperforms the cyclic code by about $5$ dB and the circulant code by more than $25$ dB. It is worthwhile to note that for rate=$1$ bpcu, the decoding search space for the proposed code is only $4$ whereas it is $256$ for the other two codes. Similarly for rate=$1.5$ bpcu, the decoding search space for the proposed code is only $8$ whereas it is $4096$ for the other two codes. Because of the exponential growth of decoding search space with increasing transmission rate for the codes in \cite{JiJ,OgH2}, it takes a prohibitively large time to complete a error performance simulation for higher rates.
\section{Discussion}
\label{sec6}

We have thus constructed a class of four group decodable DDSTCs for any power of two number of relays using algebraic techniques. It is important to note that relaxing the unitary matrix codebook to scaled unitary matrix codebook has aided us in obtaining decoding complexity benefits. A limitation of this algebraic method is that it is available only for power of two number of relays. Extending this method to obtain $g$-group decodable DDSTCs for $g>4$ appears to be difficult for two reasons:
\begin{enumerate}
\item The rate of the known linear designs in the literature for $g>4$ fall below $1$ and obtaining $R$ unitary relay matrices satisfying Condition \eqref{cond_DDSTC4} is difficult and it is not clear to the authors whether such matrices exist at all. For example, constructing the relay matrices satisfying Condition \eqref{cond_DDSTC4} for orthogonal designs is difficult.  
\item Even if the relay matrices are obtained, an initial vector satisfying Condition \eqref{cond_DDSTC5} may not exist for the chosen relay matrices.
\end{enumerate}
For the specific case of $g=4$, things were possible because the linear designs correspond to left regular matrix representation of an algebra over $\mathbb{C}$ and hence using algebraic techniques, it was possible to construct the relay matrices and the initial vector satisfying all the required conditions. 

Optimizing the signal sets for coding gain might be possible for small number of relays and for small number of points. However it becomes difficult for larger number of relays and/or larger number of points because it involves many parameters and a general closed form solution is difficult. Hence optimizing the signal sets for coding gain is an important direction for further work. 

\appendices
\section{Construction using Extended Clifford Algebra}
\label{appendix1}
\begin{defn}
Let $L=2^a,a\in \mathbb{N}$. An Extended Clifford algebra denoted by $\mathbb{A}_n^L$ is the associative algebra over $\mathbb{R}$~ generated by $n+a$ objects $\gamma_k,\ k=1,\dots,n$ and $\delta_i,\ i=1,\dots,a$ which satisfy the following relations:
\begin{itemize}
\item $\gamma_k^2=-1,~\forall~k=1,\dots,n$;~~$\gamma_k\gamma_j=-\gamma_j\gamma_k,~\forall~k\neq j$
\item $\delta_k^2=1,~\forall~k=1,\dots,a$;~~$\delta_k\delta_j=\delta_j\delta_k,~\forall~ 1\leq k,j\leq a$
\item $\delta_k\gamma_j=\gamma_j\delta_k,~\forall~1\leq k\leq a, 1\leq j\leq n$.
\end{itemize}
\end{defn}
The classical Clifford algebra, denoted by $Cliff_n$, is obtained when only the first two relations are satisfied and there are no $\delta_i$. Thus $Cliff_n$ is a sub-algebra of $\mathbb{A}_n^L$. Let $\mathscr{B}_n$ be the natural $\mathbb{R}$~ basis for this sub-algebra. 
$$
\begin{array}{rl}
\mathscr{B}_n^L=&\mathscr{B}_n\cup\left\{\mathscr{B}_n\delta_i|i=1,\dots,a\right\}\\
&\bigcup_{m=2}^{a}\mathscr{B}_n\left\{\prod_{i=1}^{m}\delta_{k_i}|1\leq k_i\leq k_{i+1}\leq a\right\}
\end{array}
$$
Then a natural $\mathbb{R}$~ basis for $\mathbb{A}_n^L$ is:\\
$$
\begin{array}{rl}
\mathscr{B}_n=&\left\{1\right\}\bigcup\left\{\gamma_i|i=1,\dots,n\right\}\\
&\bigcup_{m=2}^{n}\left\{\prod_{i=1}^{m}\gamma_{k_i}|1\leq k_i\leq k_{i+1}\leq n\right\}.
\end{array}
$$
We need a unitary matrix representation for the symbols $1$, $\gamma_1$, $\gamma_2$, $\gamma_1\gamma_2$, \mbox{$\delta_k,~k=1,\dots,a$}, \mbox{$\bigcup_{m=2}^{a}\prod_{i=1}^{m}\delta_{k_i}|1\leq k_i\leq k_{i+1}\leq a$} in the algebra $\mathbb{A}_2^L$. Such matrices are naturally provided by the left regular representation. 

We first view $\mathbb{A}_2^L$ as a vector space over $\mathbb{C}$~ by thinking of $\gamma_1$ as the complex number $i=\sqrt{-1}$. A natural $\mathbb{C}$~ basis for $\mathbb{A}_2^L$ is given by: 
$$
\begin{array}{rl}
\mathcal{B}_n^L=&\left\{1,\gamma_2\right\}\cup\left\{\left\{1,\gamma_2\right\}\delta_i|i=1,\dots,a\right\}\\
&\bigcup_{m=2}^{a}\left\{1,\gamma_2\right\}\left\{\prod_{i=1}^{m}\delta_{k_i}|1\leq k_i\leq k_{i+1}\leq a\right\}.
\end{array}
$$
Thus the dimension of $\mathbb{A}_2^L$ seen as a vector space over $\mathbb{C}$~ is $2^{n+a-1}$. We have a natural embedding of  $\mathbb{A}_2^L$ into $\mathrm{End}_{\mathbb{C}}(\mathbb{A}_2^L),$ (the set of all $\mathbb{C}$-linear maps from $\mathbb{A}_2^L$ to itself) given by left multiplication as shown below.
\begin{equation*}
\begin{array}{l}
\phi:\mathbb{A}_2^L\mapsto \mathrm{End}_{\mathbb{C}}(\mathbb{A}_2^L)\\
\phi(x)=L_x:y\mapsto xy.
\end{array}
\end{equation*}
Since $L_x$ is $\mathbb{C}$-linear, we can get a matrix representation of $L_x$ with respect to the natural $\mathbb{C}$~ basis $\mathcal{B}_n^L$. Left regular representation yields unitary matrix representations for the required symbols in the algebra. The resulting linear designs are precisely those given by Construction \ref{cons_design}. This has been explicitly shown in more detail in \cite{RaR2}. 

\section{Proof of Theorem \ref{thm_relay}}
\label{appendix2}
\begin{IEEEproof}
Let us prove the theorem assuming that all the complex variables of the linear design take values from the entire complex field. We use the fact that the linear design for $R=2^\lambda$ relays was obtained as a matrix representation of the Extended Clifford algebra $\mathbb{A}_2^{2^{\lambda-1}}$. Thus $a=\lambda-1$. We choose $M=2^a=\frac{R}{2}$. The $M$ relay matrices are explicitly given by the union of the elements of the sets $\left\{\phi(1),\phi(\delta_1),\dots,\phi(\delta_a)\right\}$ and $\left\{\bigcup_{m=2}^{a}\prod_{i=1}^{m}\phi(\delta_{k_i})|1\leq k_i\leq k_{i+1}\leq a\right\}$. By virtue of the property that $\phi$ is a ring homomorphism, these matrices are guaranteed to commute with all the codewords because they are matrix representations of elements belonging to the center of the algebra $\mathbb{A}_2^{2^{\lambda-1}}$. To obtain the remaining $R-M$ relay matrices, we need to find unitary matrices which satisfy
\begin{equation}
\label{eqn_skewcommute}
\mathbf{A_iC}^*=\mathbf{CA_i},i=M+1,\dots,R
\end{equation}
\noindent where, $\mathbf{C}$ is any codeword. But the codeword $\mathbf{C}$ is simply a matrix representation of some element belonging to the Extended Clifford algebra. One method to get these relay matrices is to take them from within the Extended Clifford algebra itself. By doing so, we can translate the condition in \eqref{eqn_skewcommute} into a condition on elements of the algebra which will then provide us a handle on the problem. Towards that end, we first identify a map in the algebra which is the analogue of taking the conjugate of the matrix representation of an element. Note from Appendix \ref{appendix1} that we used the fact that $\gamma_1$ can be thought of as the complex number $i=\sqrt{-1}$. When we take the conjugate of a matrix, we simply replace $i$ by $-i$. Hence the analogue of this action in the algebra is to replace $\gamma_1$ by $-\gamma_1$. Thus, the analogous map $\sigma$ in the algebra is defined as follows:
\begin{equation}
\sigma:x\mapsto\bar{x}
\end{equation}
where, the element $\bar{x}$ is obtained from $x$ by simply replacing $\gamma_1$ by $-\gamma_1$ in the expression of $x$ in terms of the natural ${\mathbb R}$-basis of Extended Clifford algebra. Now the problem is to find $R-M$ distinct elements denoted by $a_i,~i=M+1,\dots,R$ of the algebra $\mathbb{A}_2^{2^{\lambda-1}}$ which satisfy $a_i\bar{x}=xa_i,~\forall~x\in\mathbb{A}_2^{2^{\lambda-1}}$. The elements of the union of the following two sets satisfy the above required condition.

{\small
$$
\left\{\gamma_2\left\{1,\delta_1,\dots,\delta_a\right\}\right\}, \left\{\gamma_2\left\{\bigcup_{m=2}^{a}\prod_{i=1}^{m}\delta_{k_i}|1\leq k_i\leq k_{i+1}\leq a\right\}\right\}
$$ 
}
This can be proved by using the anti-commuting property, i.e., $\gamma_2(-\gamma_1)=\gamma_1(\gamma_2)$. Hence the matrix representation of these specific elements gives the unitary relay matrices $\mathbf{A_i},i=M+1,\dots,R$. Suppose we plug in these relay matrices to form a linear design \mbox{$\mathbf{X}=\left[\begin{array}{cccccc}\mathbf{A_1s} & \dots & \mathbf{A_Ms} & \mathbf{A_{M+1}s}^* & \dots & \mathbf{A_Rs}^*\end{array}\right]$} where, $\mathbf{s}=\left[\begin{array}{cccc}x_1 & x_2 & \dots & x_R\end{array}\right]^T$, it turns out that we get exactly the same linear design which is used at the source. Thus the initial vector choice of $\mathbf{s_0}=\left[\begin{array}{cccc}1 & 0 & \dots & 0\end{array}\right]^T$ guarantees that the initial matrix $\mathbf{X_0}$ is an identity matrix.

However, we would like to point out that there are also other elements of the algebra which satisfy \mbox{$a_i\bar{x}=xa_i,~\forall~x\in\mathbb{A}_2^{2^{\lambda-1}}$}. For example, consider the union of the elements of the sets $\left\{\gamma_1\gamma_2\left\{1,\delta_1,\dots,\delta_a\right\}\right\}$ and $\left\{\gamma_1\gamma_2\left\{\bigcup_{m=2}^{a}\prod_{i=1}^{m}\phi(\delta_{k_i})|1\leq k_i\leq k_{i+1}\leq a\right\}\right\}$.
\end{IEEEproof}

\section*{Acknowledgment}
The authors thank Prof. Hamid Jafarkhani, Dr. Yindi Jing and Dr. Fr\'{e}d\'{e}rique Oggier for providing us with preprints of their recent works \cite{JiJ,OgH1,OgH2,Ogg}. The authors are grateful to the anonymous reviewers for providing constructive comments which helped in improving the presentation of this paper.

\ifCLASSOPTIONcaptionsoff
  \newpage
\fi




\begin{thebibliography}{1}

\bibitem{JiH1} Y. Jing and B. Hassibi, ``Distributed space-time coding in wireless relay networks," \emph{IEEE Trans. Wireless Commun.}, vol. 5, no. 12, pp. 3524-3536, Dec 2006.

\bibitem{KiR} Kiran T. and B. Sundar Rajan, ``Partially-coherent distributed space-time codes with differential encoder and decoder,'' \emph{IEEE J. Select. Areas Commun.}, vol. 25, no. 2, pp. 426-433, Feb 2007.

\bibitem{JiJ} Y. Jing and H. Jafarkhani,``Distributed differential space-time coding for wireless relay networks," to appear in \emph{IEEE Trans. on Commun.}. Private Communication.

\bibitem{OgH1}  F. Oggier, B. Hassibi, "A coding strategy for wireless networks with no channel informations", in \emph{Allerton}, 2006.

\bibitem{OgH2}  F. Oggier, B. Hassibi, "Cyclic distributed space-time codes for wireless relay networks with no channel information,'' submitted for publication. 

\bibitem{TaC} M. Tao and R. S. Cheng, ``Differential space-time block codes," Proceedings of \emph{IEEE Globecom 2001}, vol. 2, pp. 1098-1102, San Antonio, USA, Nov 25-29, 2001.

\bibitem{YGT2} Chau Yuen, Yong Liang Guan, Tjeng Thiang Tjhung, ``Single-symbol decodable differential space-time modulation based on QO-STBC," \emph{IEEE Trans. Wireless Commun.}, vol. 5, no. 12, pp. 3329-3335, Dec 2006. 

\bibitem{TBH} O. Tirkkonen, A. Boariu and A. Hottinen, ``Minimal non-orthogonality rate 1 space-time block code for $3+$ tx antennas,'' in \emph{IEEE Int. Symp. on Spread-Spectrum Tech. \& Appl.}, New Jersey, Sep 6-8, 2000, pp. 429-432.  

\bibitem{Ogg} F. Oggier, ``Cyclic algebras for noncoherent differential space-time coding," \emph{IEEE Trans. Inform. Theory}, vol. 53, no. 9, pp. 3053-3065, Sep 2007.

\bibitem{RaR2} G. Susinder Rajan and B. Sundar Rajan, ``Algebraic distributed space-time codes with low ML decoding complexity,'' Proceedings of \emph{IEEE Intl. Symp. Inform. Theory}, Nice, France, June 24-29, 2007, pp. 1516-1520.

\bibitem{RaR4} --------, ``Noncoherent low-decoding-complexity space-time codes for wireless relay networks,''in \emph{IEEE Intl. Symp. Inform. Theory}, Nice, France, June 24-29, 2007, pp. 1521-1525.

\bibitem{RaR5} --------, ``Signal set design for full-diversity low-decoding complexity differential scaled-unitary STBCs,'' in \emph{IEEE Intl. Symp. Inform. Theory}, Nice, France, June 24-29, 2007, pp. 1616-1620.


\end{thebibliography}
%

%

\begin{IEEEbiography}[{\includegraphics[width=1in,height=1.25in,clip,keepaspectratio]{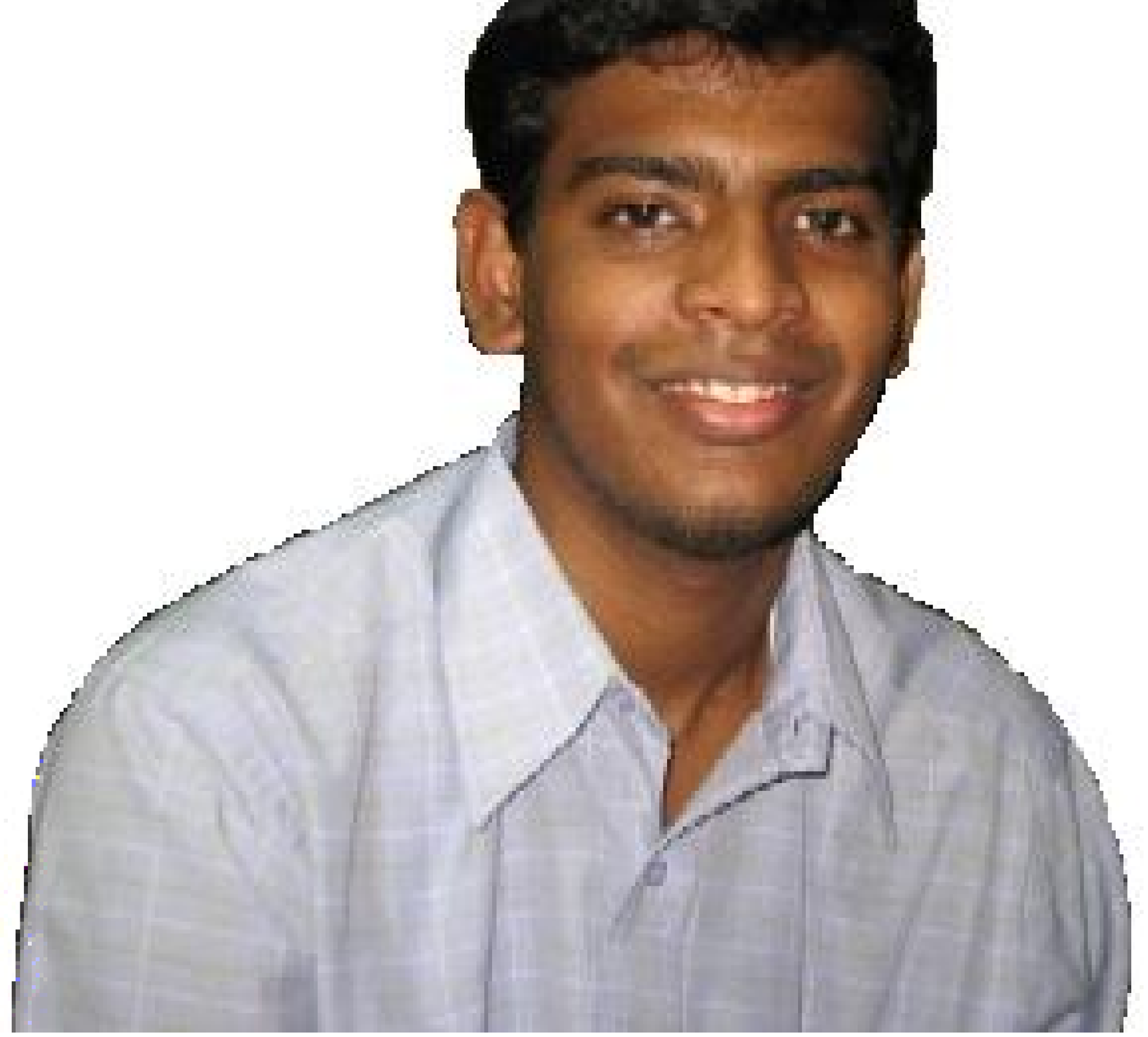}}]{G. Susinder Rajan}
(S'2006) was born in Chennai, India in 1983. He completed his B.E. degree at the College of Engineering Guindy, Anna University, India in 2005. He is currently a Ph.D. student in the Department of Electrical Communication Engineering, Indian Institute of Science, Bangalore, India. His primary research interests include space-time coding for MIMO channels and wireless relay channels with an emphasis on algebraic code construction techniques. 
\end{IEEEbiography}

\vfill

\begin{IEEEbiography}[{\includegraphics[width=1in,height=1.25in,clip,keepaspectratio]{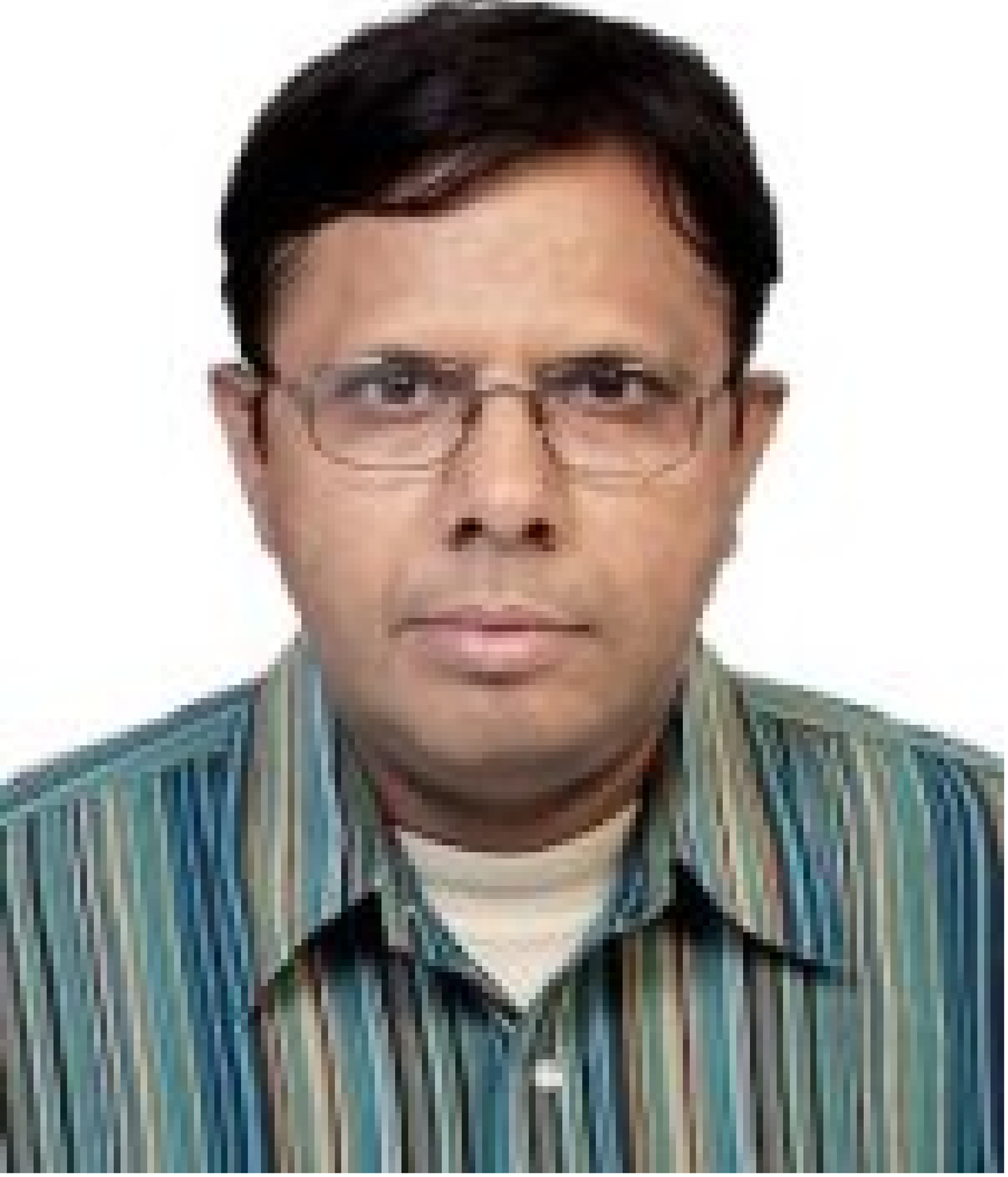}}]{B. Sundar Rajan}
(S'84-M'91-SM'98) was born in Tamil Nadu, India. He received the B.Sc. degree in mathematics from Madras University, Madras, India, the B.Tech degree in electronics from Madras Institute of Technology, Madras, and the M.Tech and Ph.D. degrees in electrical engineering from the Indian Institute of Technology, Kanpur, India, in 1979, 1982, 1984, and 1989 respectively. He was a faculty member with the Department of Electrical Engineering at the Indian Institute of Technology in Delhi, India, from 1990 to 1997. Since 1998, he has been a Professor in the Department of Electrical Communication Engineering at the Indian Institute of Science, Bangalore, India. His primary research interests are in algebraic coding, coded modulation and space-time coding.

Dr. Rajan is an Editor of IEEE Transactions on Wireless Communications from 2007 and also a Editorial Board Member of International Journal of Information and Coding Theory. He is a Fellow of Indian National Academy of Engineering and  is a Member of the American Mathematical Society. 
\end{IEEEbiography}
\vfill

\end{document}